\documentstyle[12pt]{article}

\begin{document}
\title{PLANCKSUSY - new program for SUSY masses calculations: from Planck 
scale to our reality}
\author{N.V.Krasnikov\thanks{email address : Nikolai.Krasnikov@cern.ch}
 and V.V.Popov\thanks{email address : Vsevolod.Popov@cern.ch}
\\Institute for Nuclear Research, Moscow 117312\\}

\date{November, 1996}
\maketitle
\begin{abstract}
We describe briefly new program for SUSY masses calculations. The main 
distinction of our program is that we start to solve the renormalization 
group equations for soft SUSY  breaking parameters for SU(5) SUSY GUT 
model from Planck scale $M_{PL} = 2.4\cdot 10^{18}$ Gev. Our program 
works also for large $\tan(\beta)$. We find that for 
$m_{0} \leq 0.5\cdot m_{\frac{1}{2}}$ the effects of the evolution from 
Planck scale to GUT scale are very essential. In particular, we find that 
neutralino even for small $m_{0}$ is always LSP. We also introduced in our 
program some parameter of non-universality of the gaugino masses at GUT scale. 
Playing with the non-universality of the gaugino masses at GUT scale it is 
possible to have the situation when leptonic modes are suppressed and the 
single SUSY signature is hadronic jets with missing energy.

\end{abstract}

\begin{flushright}
Preprint INR 976TH/96

\end{flushright}     

\newpage

Supersymmetric electroweak models offer the simplest solution of the 
gauge hierarchy problem \cite{1}-\cite{4}. In real life supersymmetry has 
to be broken and the masses of superparticles have to be lighter than 
$O(1)$ Tev  \cite{4}. Supergravity gives natural explanation of the 
supersymmetry breaking, namely, an account of the supergravity breaking in 
hidden sector leads to soft supersymmetry breaking in observable sector 
\cite{4}. For the supersymmetric extension of the Weinberg-Salam model 
soft supersymmetry breaking terms usually consist of the gaugino mass 
terms, squark and slepton mass terms with the same mass at GUT scale and 
trilinear soft scalar terms proportional to the superpotential \cite{4}.

In this paper we briefly describe our program "PLANCKSUSY" for SUSY masses 
calculations ( the program can be received by e mail: Vsevolod.Popov@cern.ch).
As it has been mentioned in the abstract the main peculiarity of the program is 
that we can start not from GUT scale $M_{GUT} \approx 2\cdot 10^{16}$Gev but 
also from Planck scale $M_{PL} = 2.4 \cdot 10^{18}$ Gev. We assume that physics 
between Planck scale and GUT scale is described by standard supersymmetric 
SU(5) GUT model \cite{4}. Of course, we don't know for sure the physics 
between Planck and GUT scales so we consider SU(5) SUSY model as the 
simplest possibility to use it as a "quasi-realistic" model to take into 
account the effects of the evolution between Planck and GUT scales. The 
standard assumption \cite{4,5} is to start from GUT scale and to impose 
universal mass relations for squark, slepton and Higgs soft masses 
and for gauginos at GUT scale 
\begin{equation}
m_{sq} = m_{sl} = m_{H} = m_{0},
\end{equation}
\begin{equation}
m_{3} = m_{2} = m_{1} = m_{\frac{1}{2}}
\end{equation}
However in standard approach soft SUSY breaking terms arise as a result of 
the supergravity breaking in hidden sector so it is more valid to impose 
universal boundary conditions not at GUT scale but at Planck scale. It should 
be noted that in superstring inspired models in general we  expect 
non-universal boundary conditions for SUSY soft breaking terms \cite{6}. 
To take into account possible effects from non-universal boundary conditions 
we introduced in our program some parameter of non-universality for gaugino 
masses at GUT scale. It should be noted that our program takes 
nonzero Yukawa couplings for t-quark, b-quark and tau-lepton so we can use 
it for large values of $\tan(\beta)$. Besides we can start both from Planck 
and GUT scales to compare the results of the calculations.

Our program can be divided roughly speaking into three  parts. At first  
stage we solve numerically renormalization group equations for SUSY soft 
breaking parameters and coupling constants for SU(5) SUSY model between 
Planck and  GUT scales. Here we  use  one loop RG equations of ref.\cite{7}.
At second stage  we solve numerically RG equations for $SU(3)\otimes 
SU(2)\otimes U(1)$ MSSM between $M_{GUT}$ and $M_{SUSY}$. We use two loop 
RG equations for couplings and gaugino masses and one loop RG equations 
for other soft SUSY breaking parameters \cite{8}. From $M_{SUSY}$ to 
$m_{t}$ we use RG equations for standard WS model. At third stage we solve 
the equations for the determination of the electroweak minimum of 
the one loop effective potential to determine the absolute value of the $\mu\/$ 
parameter and the Higgs boson masses. We also calculate mixing effects for 
neutralino, chargino and stops. We calculate the pole gluino mass and 
other running sparticle masses $m(m)$. The difference between running and 
pole masses for squarks and other sparticles is rather small. 
Here we use the formulae of refs.\cite{9}-\cite{11}. The input parameters 
of our program are the standard ones ($m_{0}, m_{\frac{1}{2}}, \tan(\beta).
sign(\mu), A,m_t\/$). Our program can start both from 
Planck scale and GUT scale. We also introduced explicitly some parameter  
of the non-universality for gaugino masses at GUT scale. We use the following 
formula for gaugino mass matrix at GUT scale:
\begin{equation}
m = m_{\frac{1}{2}} I + k\Phi,
\end{equation}       
where I is the unit matrix and $\Phi = {\Phi}_{0}Diag(2,2,2,-3,-3)$
Parameter k determines the non-universality of gaugino masses at GUT scale. 
Our formulae for SU(3), SU(2) and U(1) gaugino masses can be written in 
the form 
\begin{equation} 
m_{3} = m_{\frac{1}{2}}(1 + 2\delta)  ,
\end{equation}
\begin{equation}
m_{2} = m_{\frac{1}{2}}(1 -  3\delta) ,
\end{equation}
\begin{equation}
m_{1} = m_{\frac{1}{2}}(1 - \delta)
\end{equation}
Here $\delta$ is some parameter of the gaugino mass non-universality. 

Let us briefly consider first nontrivial physical consequences of our program. 
We have found that for  $m_{0} \geq 0.5\cdot m_{\frac{1}{2}}$ both 
"Planck" and "GUT" spectra coincide up to 20 percent. For $m_{0} \leq 
  0.5\cdot m_{\frac{1}{2}}$ we find essential difference in slepton spectrum. 
The most interesting difference is that for  small $m_{0}$ and $m_{\frac{1}{2}} 
\geq O(300)$ Gev the lightest superparticle in "GUT" model is charged  
right-handed tau slepton that contradicts to the  experimental data on 
abundances of anomalous super-heavy isotopes and usually this region in $(m_{0}, 
m_{\frac{1}{2}}$ plane  is  considered to be "theoretically excluded".  
When we start from Planck scale the lightest superparticle 
in this region of $(m_{0}, m_{\frac{1}{2}})$ parameters is always neutralino 
so we don't have at all "excluded" region. Moreover, it 
appears that it is general situation when we start not from GUT scale but from 
Planck scale (that  is more "scientific"). Even if we work within 
$SU(3) \otimes SU(2) \otimes U(1)$ MSSM and  impose boundary conditions 
at Planck scale we find that in this case also the lightest sparticle is 
neutralino. For instance, for $m_{0} =0 , m_{\frac{1}{2}} = 500 Gev,  A =  0,  
sign(\mu) = - $ we find that in "PLANCK"("GUT") model the masses are:\\
 $m(\tilde{g})=1377(1278)$ Gev,\hspace{3cm}
 $m(\tilde{u}_{L}) = 1242(1118)$ Gev,\\
 $m(\tilde{d}_{L}) = 1244(1120)$ Gev,\hspace{3cm}
  $m(\tilde{u}_{R}) = 1199(1077)$ Gev,\\
$m(\tilde{b}_{L}) = 1115(1019)$ Gev,\hspace{3cm}
 $m(\tilde{b}_{R}) = 1177(1072)$ Gev,\\
$m(\tilde{t1} = 918(841)$ Gev,\hspace{3cm}
 $m(\tilde{t2}) = 1120(1027)$ Gev,\\
$m(\tilde{\nu}_{L}) = 471(348)$ Gev,\hspace{3cm}
  $m(\tilde{e}_{L} = 475(353)$ Gev,\\
$m(\tilde{e}_{R})  = 404(196)$ Gev,\hspace{3cm}
 $m(\tilde{\nu_{\tau_{L}}}) = 471(348)$ Gev,\\
$m(\tilde{\tau}_{L}) = 475(352)$ Gev,\hspace{3cm}
 $m(\tilde{\tau_{R}})  = 364(196)$ Gev,\\
$m(\chi_1) = 226(205)$ Gev,\hspace{3cm}
 $m(\chi_2) = 441(401)$ Gev,\\
 $m(\chi_3) = 989(886)$ Gev,\hspace{3cm}
 $m(\chi_4) = 993(890)$ Gev,\\
 $m(\chi_{1}^{+}) = 441(401)$ Gev ,\hspace{3cm}
$m(\chi_{2}^{+}) = 993(890)$ Gev,\\
 $m(h) =  108(106)$ Gev,\hspace{3cm}
 $m(H^{o}) = 1255(1102)$ Gev,\\
 $m(A) = 1252(1098)$ Gev,\hspace{3cm}
 $m(H^{+}) = 1254(1101)$ Gev.\\
We think that irrespective of validness of our concrete SU(5) model for 
the description of physics between Planck and GUT scales for small 
$m_{0}$ the effects of the evolution from  Planck to GUT scales are 
very important. Note also that for "Planck" model we have rather big splitting 
among the sleptons of the first two generations and the third one (in our 
program sparticle masses of the first and second generations coincide).
Second lesson from our program is that for the case of  strong violation of 
gaugino mass universality at GUT scale we can have the situation when 
the second neutralino  and the first chargino  being mainly SU(2) gaugino 
are heavier than the gluino. For instance, for $m_{0} =800 $Gev, 
$m_{\frac{1}{2}} = 200$ Gev, $A =0$, $\tan{\beta} = 2$, $\delta = -1$, 
$sign(\mu) = - $ in "GUT" model gluino mass is $m(\tilde{g}) = 599$ Gev, 
whereas the neutralino and chargino made basically from gaugino 
with small mixing of higgsino have the following masses:

$m(\chi_{1}^0) = 168$ Gev, $m(\chi_{2}^0) = 664$ Gev, 
$m(\chi_{1}^{+}) = 664$ Gev.  
All squarks masses here are heavier than the gluino mass. 
It means that gluino can decay only to quark-antiquark and  LSP  so we 
shall not  have  multilepton signatures and the single SUSY signature at LHC  
and TEVATRON  will be multijets with missing $E_{T}$. Moreover for the case  
of the  model with R-parity violation of the $uds$ type LSP will decay into two 
quarks that makes the observation of such particular scenario very  
difficult at LHC.

\section*{A brief overview of the code}
A code is written on FORTRAN 77; we might not be keeping to all the 
standards, but at least, it runs OK under UNIX on CERN clusters.
Double precision is used throughout
all the calculations. Systems of algebraic and differential equations 
are solved by calls to CERNLIB \cite{CL} subroutines DSNLEQ and DDEQMR.
For neutralino matrix eigenvalues subroutine RSM is used
 (taken from \cite{subr}). Some checks are performed during the calculations,
so for some ``bad'' initial values,  the code
skips further calculations at certain moment and returns with non-zero 
error flag (see below). Still, no checks on any physical meaning is made, 
 so it is up to user to analyse output masses and error flag.
Essentially procedure consists of solving the systems of differential 
equations with some of the variables ($h_t,h_\tau \/$) being fixed
in fact at the final scale. To find their values at the  starting scale
( $M_{GUT}\/$ or $M_{Planck}\/$) we start with the solving the system 
of equations, decoupled  indeed  for most (but not for all!) reasonable input
parameters :
\begin{eqnarray}
\nonumber h_t(x) \mid _{electroweak\hspace{6pt}scale} = ``fixed''h_t,\\
\nonumber h_\tau(y) \mid _{electroweak\hspace{6pt}scale} = ``fixed''h_\tau
\end{eqnarray}
 in respect to  $x,y\/$ (values of $h_t,h_\tau\/$ at starting scale).
Their  values at  electroweak scale are fixed by the values of $m_t,m_\tau,
\alpha_s\/$. In the process of solution, we find also the scale, taken as
$\sqrt{m(\tilde {t_1})m(\tilde {t_2})}\/$ to change from two- to one-loop
RG equations. To get the values of $m(\tilde {t_1}),m(\tilde {t_2})\/$ we
solve the equation  to find electroweak minimum  $\mu\/$.
 The rest of program  flow is rather trivial arithmetics except for masses
``tuning'' (their values are taken not at common scale but each one
is calculated at the scale equal to this very mass). It requires again the
solution of equation for the value of the mass, now with the final  scale
taken as variable.
\newline Masses calculation is performed by the call to steering subroutine
SUSYM with the calling sequence : 
\begin{verbatim}
      call susym
     $(iway,AM0,AM_HALF,A_t,tanb,AMT,csign_mu,isec,dnu,idebug,
     $amass,ierr)
\end{verbatim}
User must declare 
\begin{verbatim}
      integer iway(3)
      real amass(42)
      character*1 csign_mu
\end{verbatim}
Input parameters :
\begin{itemize}
\item
\begin{itemize}
\item
iway(1) = 1 : to run from Planck to weak scale
\item
iway(1) = 2 : from GUT to weak scale
\item
iway(2) = 1 : normal run ( all calculations of all the masses)
\item
iway(2) = 2 : Stop after $h_t\/$, $h_b\/$ tuning. Useful in looking for
possible pairs of ($tan(\beta)\/$,$M_t\/$)
\item
 iway(3) is not used for now
\end{itemize}
\item
 AM0 : $M_0\/$ at initial scale (real)
\item
 AM\_HALF : $M_{1/2}\/$ at initial scale (real)
\item
 A\_t : $A_t\/$(real)
\item
 tanb : $tan(\beta)\/$ (real)
\item
 AMT  : $M_t\/$, mass of top-quark (real)
\item
 isec : second loop coefficient (integer); set it normally to 1
\item
 dnu : non-universality coefficient (real)
\item
 idebug :  produce some output if $> 0\/$ (integer) 
\end{itemize}
Output parameters :\\
\begin{itemize}
\item
array amass : values/masses at the weak scale
\begin{enumerate}
\item       $g^2_1$ 
\item       $g^2_2$ 
\item       $g^2_3$ 
       
\item       $M_{1/2}(1)$ 
\item       $M_{1/2}(2)$ 
\item       $M_{1/2}(3)$ 
       
\item       $h_t$ 
\item       $h_b$ 
\item       $h_\tau$ 
       
\item       $A_t$
\item       $A_b$ 
\item       $A_\tau$ 
       
\item       $\tilde U_l$ 
\item       $\tilde D_l$ 
\item       $\tilde U_r$ 
\item       $\tilde D_r$ 
       
\item       $\tilde \nu_l$ 
\item       $\tilde e_l$ 
\item       $\tilde e _r$ 
       
\item       $\tilde B_l$ 
\item       $\tilde T_l$ 
\item       $\tilde B_r$ 
\item       $\tilde T_r$ 
       
\item       $\tilde {\nu_{\tau_l}}$ 
\item       $\tilde \tau_l$ 
\item       $\tilde \tau_r$ 
       
\item       $\mu$ 
       
\item       $\tilde T_1$ 
\item       $\tilde T_2$ 
\item       $\tilde H_1$ 
\item       $\tilde H_2$ 
       
\item       $\chi_1^0$
\item       $\chi_2^0$
\item       $\chi_3^0$
\item       $\chi_4^0$
       
\item       $\chi_1^+$
\item       $\chi_2^+$
       
\item       $\tilde H_{axial}$  
\item       $\tilde H_{charged}$  
\item       $\tilde H_{light}$  
\item       $\tilde H_{heavy}$  
       
\item       Gluino 
\end{enumerate}
\item
      ierr = 
\begin{itemize}
\item            0     OK
\item            1     error in input parameters
\item           10     bad fitting of $h_t,h_b\/$
\item          100     negative $m^2\/$ for some of sparticles
\item         1000     first approximation for ${\mu}^2\/$ is negative
\item         2000     low and high limits for ${\mu}^2\/$  are 0. both
\item         3000     no good solution for $\mu\/$ found
\item        10000     failed to find eigenvalues for neutralinos
\item       100000     failed to find eigenvalues for charginos
\end{itemize}
\end{itemize}

To conclude, we have written "PLANCKSUSY" program for SUSY masses  
calculations. The main peculiarity of our program 
for instance in comparison with ISASUSY program \cite{12} is that we can 
start from both Planck and GUT scales. We also introduced the non-universality  
parameter for gaugino masses. It appears that for very interesting region 
in the parameter space when $m_{0}$ is much smaller than 
$m_{\frac{1}{2}}$ if we start from the Planck scale we find  that  in this 
region the LSP is always neutralino unlike to the case when we start from 
GUT scale and LSP is the right-handed tau-slepton. For the case when we have 
strong deviation from gaugino mass universality it is possible to realize the 
situation when the second neutralino and the first chargino are heavier than 
gluino. For such scenario lepton signatures for SUSY search  at LHC and 
TEVATRON are absent and the single SUSY signature are multijet events with 
missing $E_{T}$. Moreover, for the models with R-parity violation of $uds$ 
type we shall have only jet events in final states that makes SUSY 
discovery for such scenario rather difficult. The program can be received 
by e mail:Vsevolod.Popov@cern.ch     
           
We thank CERN CMS Department for the hospitality during our  stay at CERN where 
this paper has been finished. We are  indebted to the participants of the 
Daniel Denegry CMS physics simulation seminar and espessially to Daniel for 
vivid discussions. We also indebted  to I.Semenjuk for  the help and  
consultations.

\end{document}